\documentclass[conference]{IEEEtran}
\usepackage{graphicx}
\usepackage{amssymb}
\usepackage{bm}
\usepackage{amsmath}
\usepackage{pgfplots}
\usepackage{epstopdf}
\pgfplotsset{compat=newest}
\pgfplotsset{plot coordinates/math parser=false}
\newlength\figureheight
\newlength\figurewidth

\usepackage{color,soul}
\newcounter{MYtempeqncnt}
\newcommand\ccounter{16}
\IEEEoverridecommandlockouts

\begin{document}

\title{On the Performance of Cell-Free Massive MIMO with Short-Term Power Constraints}
\author{\IEEEauthorblockN{Giovanni Interdonato$^{*\dagger}$, Hien Quoc Ngo$^\dagger$, Erik G. Larsson$^\dagger$, P{\aa}l Frenger$^*$}
\IEEEauthorblockA{$^*$Ericsson Research, Wireless Access Networks, 581 12 Link\"oping, Sweden\\
$^\dagger$Department of Electrical Engineering (ISY), Link\"oping University, 581 83 Link\"oping, Sweden\\
\{giovanni.interdonato, pal.frenger\}@ericsson.com, \{hien.ngo, erik.g.larsson\}@liu.se\\}
\thanks{This paper was supported by the European Union's Horizon 2020 research
and innovation programme under grant agreement No 641985 (5Gwireless).}}
\maketitle

\begin{abstract}
In this paper we consider a time-division duplex cell-free massive multiple-input multiple-output (MIMO) system where many distributed access points (APs) simultaneously serve many users. A normalized conjugate beamforming scheme, which satisfies short-term average power constraints at the APs, is proposed and analyzed taking into account the effect of imperfect channel information. We derive an approximate closed-form expression for the per-user achievable downlink rate of this scheme. We also provide, analytically and numerically, a performance comparison between the normalized conjugate beamforming  and the conventional conjugate beamforming scheme in \cite{NgoCF} (which satisfies long-term average power constraints).
Normalized conjugate beamforming scheme reduces the beamforming uncertainty gain, which comes from the users' lack of the channel state information knowledge, and hence, it improves the achievable downlink rate compared to the conventional conjugate beamforming scheme.
\end{abstract}


\section{Introduction}

Cell-free massive multiple-input multiple-output (MIMO), also known as distributed massive MIMO \cite{NgoCF}, \cite{ZhouWCS}, \cite{Truong}, has the characteristics of a massive MIMO system \cite{MarzettaNonCooperative} where the antennas are spread out over a large area, in a well-planned or random fashion. These antennas, called access points (APs) herein, simultaneously serve many users in the same frequency band. In cell-free massive MIMO, all APs are involved in coherently serving all users. Due to the network topology, cell-free massive MIMO has great abilities to spatially multiplex users and to control the interference \cite{NgoCF}. The increased macro-diversity gain leads to improved coverage probability and energy efficiency compared to co-located massive MIMO \cite{Kamga}. This comes at the price of higher backhaul requirements and the need for distributed signal processing. Hence, cell-free massive MIMO has attracted a lot of research interest recently.

When determining the power allocation to be used in a cell-free massive MIMO system (and in general for every wireless system) we need to take into account that power is constrained by either long-term average power constraints or short-term average power constraints \cite{Khoshnevisan}. For long-term power constraints, the average is taken over codewords and channel fading coefficients. By contrast, for short-term power constraints, the average is just taken over the codewords. Paper  \cite{NgoCF} consider cell-free massive MIMO systems with conjugate beamforming precoding subject to long-term average power constraints. To the best of the authors' knowledge, no analysis on cell-free massive MIMO subject to short-term average power constraints is available in literature.

\textbf{Contributions:} We propose a downlink precoding scheme, named as \textit{normalized conjugate beamforming}, that satisfies short-term average power constraint at the APs, and derive an approximate closed-form expression for the per-user achievable downlink rate of this scheme.  Our analysis takes into account channel estimation errors and the effect of adopting arbitrary uplink pilots, i.e., \textit{pilot contamination}. Numerical results verify the tightness of our approximation. We further compare the performance of the normalized conjugate beamforming with the conventional conjugate beamforming  in  \cite{NgoCF}. Normalized conjugate beamforming hardens the effective channel gains at the users, and hence, outperforms the conventional conjugate beamforming when the number of APs is moderate.       


\section{System Model and Notation}

Consider a time-division duplex (TDD) cell-free massive MIMO system, in which $M$ single-antenna APs, distributed in a wide area, simultaneously serve $K$ single-antenna users in the same frequency band, and $M>K$. The $K$ users are also located at random in the same area. We assume the hardware calibration is perfect so that the channels are reciprocal. Reciprocity calibration, to the required accuracy, can be practically achieved using off-the-shelf methods \cite{Lund}.  
A backhaul network connects all the APs with a central processing unit (CPU), which is responsible of exchanging information such as payload data and power control coefficients. Channel state information (CSI) acquisition and precoding are carried out locally at each AP.

The time-frequency resources are divided into coherence intervals of length $\tau$ symbols. Coherence interval is defined as the interval during which the channel is approximately constant. Therefore, we assume that channel is static within a coherence interval and varies independently between every coherence interval.  

Let $g_{mk}$ be the channel coefficient between the $k$th user
and the $m$th AP, defined as
\begin{equation}
\label{eq:channelcoeff}
g_{mk} = \sqrt{\beta_{mk}}h_{mk},
\end{equation}
for $m=1,\ldots,M$, and $k=1,\ldots,K$, where $h_{mk}$ is the small-scale fading, and $\beta_{mk}$ represents the large-scale fading. In cell-free massive MIMO, the large-scale fading coefficients $\{\beta_{mk}\}$ depend on both $m$ and $k$, since the APs are spread out in a large area and not co-located as in conventional massive MIMO systems. They are constant for several coherence intervals. Furthermore, we assume that the coefficients $\{\beta_{mk}\}$ are estimated a-priori, and known, whenever required. The small-scale fading is modeled as Rayleigh fading, i.e. the coefficients $\{h_{mk}\}$ are i.i.d.\ $\mathcal{CN}(0,1)$ RVs.


Each coherence interval is divided into three phases: uplink training, uplink payload data transmission, and downlink payload data transmission. Since we focus on the downlink performance of cell-free massive MIMO, the performance analysis of the uplink payload data transmission phase is omitted.

\subsection{Uplink Training and Channel Estimation}

In the uplink training phase, users synchronously send pilots to all the APs. Then, based on the received pilot signals, each AP estimates the channels to all users. The APs use the channel estimates to perform the signal detection during the uplink data transmission phase, and to precode the transmitted symbols during the downlink data transmission phase.

Let $\tau_\textrm{u,p}$ be the number of symbols per coherence interval spent on the transmission of uplink pilots such that $\tau_\textrm{u,p}<\tau$. 
The uplink pilot sequence sent by the $k$th user is denoted by $\sqrt{\tau_\textrm{u,p}}\bm{\varphi}_k \in \mathbb{C}^{\tau_\textrm{u,p}\times1}$, where $\Vert\bm{\varphi}_k\Vert^2=1$. 
Pilot sequences assigned for all users should be pairwisely orthogonal. However, the maximum number of orthogonal pilot sequences is upper-bounded by the uplink training duration $\tau_\textrm{u,p}$ which depends on the length of the coherence interval $\tau$. Adopting orthogonal pilots leads to inefficient resource allocation when the number of the users is large or when the coherence interval is short. For these cases, pilot sequences must be reused among users. As a consequence,  the channel estimate of a given user is contaminated by interference from other users which share the same pilot sequences. This is called \textit{pilot contamination} \cite{PilotContamination}. In this work, we consider a general case where the pilot sequences $\{\bm{\varphi}_k\}$ are arbitrary. 

The $m$th AP receives a linear combination of $K$ uplink pilots, i.e., a $\tau_\textrm{u,p}\times1$ vector given by
\begin{equation}
\label{eq:uplinkpilot}
\textbf{y}_{\textrm{up},m} = \sqrt{\tau_\textrm{u,p}\rho_\textrm{u,p}}\sum^K_{k=1} g_{mk}\bm{\varphi}_k + \textbf{w}_{\textrm{up},m},
\end{equation}
where $\rho_\textrm{u,p}$ is the normalized signal-to-noise ratio (SNR) of the pilot symbol and $\textbf{w}_{\textrm{up},m}$ is the additive noise vector  whose elements are i.i.d. $\mathcal{CN}(0,1)$ RVs. The received signal is processed locally at the AP side by projecting it on $\bm{\varphi}^H_k$ as
\begin{align}
\label{eq:uplinkpilotprojection}
& \check{y}_{\textrm{up},mk} = \bm{\varphi}^H_k\textbf{y}_{\textrm{up},m} \nonumber \\ 
&= \sqrt{\tau_{\textrm{u,p}}\rho_{\textrm{u,p}}} g_{mk}+ \sqrt{\tau_{\textrm{u,p}}\rho_{\textrm{u,p}}}\sum\limits^K_{k'\neq k} g_{mk'} \bm{\varphi}^H_k \bm{\varphi}_{k'} + \bm{\varphi}^H_k\textbf{w}_{\textrm{up},m}.
\end{align}
Due to the pilot contamination effect represented by the second term in (\ref{eq:uplinkpilotprojection}), performing linear \textit{minimum mean-square error} (MMSE) estimation of the channel $g_{mk}$, based on  $\check{y}_{\textrm{up},mk}$, leads to a suboptimal estimates. This is optimal only in the special case where any two pilots are orthogonal or the same. The linear MMSE estimate of $g_{mk}$ given $\check{y}_{\textrm{up},mk}$, is \cite{NgoCF}
\begin{equation}
\label{eq:mmse}
\hat{g}_{mk} = \frac{\mathbb{E}\{\check{y}^*_{\textrm{up},mk}g_{mk}\}}{\mathbb{E}\{\vert\check{y}_{\textrm{up},mk}\vert^2\}}\check{y}_{\textrm{up},mk} = c_{mk}\check{y}_{\textrm{up},mk},
\end{equation}
where
\begin{equation}
\label{eq:cmk}
c_{mk} \triangleq \frac{\sqrt{\tau_{\textrm{u,p}}\rho_{\textrm{u,p}}}\beta_{mk}}{\tau_{\textrm{u,p}}\rho_{\textrm{u,p}} \sum^K_{k'=1}\beta_{mk'}|\bm{\varphi}^H_k \bm{\varphi}_{k'}|^2+1}.
\end{equation}
The corresponding channel estimation error is defined as $\tilde{g}_{mk} \triangleq g_{mk} - \hat{g}_{mk}$. Note that, since we use linear MMSE estimation, the channel  estimate $\hat{g}_{mk}$ and the estimation error $\tilde{g}_{mk}$ are uncorrelated. Furthermore, since $g_{mk}$ is Gaussian distributed, $\hat{g}_{mk}$ in (\ref{eq:mmse}) becomes the MMSE estimate of
$g_{mk}$. As a result, $\hat{g}_{mk}$ and $\tilde{g}_{mk}$ are independent. The variance of $\hat{g}_{mk}$ is denoted by
\begin{equation}
\label{eq:defGamma}
\gamma_{mk} \triangleq \mathbb{E}\{{|\hat{g}_{mk}|}^2\} = \sqrt{\tau_\textrm{u,p}\rho_\textrm{u,p}} \beta_{mk} c_{mk}.
\end{equation}
Note that, $\gamma_{mk}$ depends on the large-scale fading and $ \gamma_{mk}\leq\beta_{mk} $. If the channel estimation is perfect, then $\gamma_{mk}=\beta_{mk}$. 

\subsection{Downlink Data Transmission}

In the downlink data transmission phase, all $M$ APs use the channel knowledge acquired during the uplink training phase and the conjugate beamforming with short-term power constraint technique to precode the signals intended for the $K$ users. More precisely, the $m$th AP transmits the data signal $x_m$, which is given by
\begin{equation}
\label{eq:DLdata}
x_m = \sqrt{\rho_\textrm{d}}\sum^K_{k=1} \sqrt{\eta_{mk}} \frac{\hat{g}^*_{mk}}{|\hat{g}_{mk}|} q_k,
\end{equation}
where $q_k$ is the data symbol intended for the $k$th user satisfying $\mathbb{E}\{\vert q_k \vert^2\}=1$, and $\rho_\textrm{d}$ is the normalized transmit SNR of the $m$th AP. Term $\hat{g}^*_{mk}/|\hat{g}_{mk}|$ is defined as \textit{precoding factor}. Herein, we named this precoding scheme as \textit{normalized conjugate beamforming}, in order to distinguish it from the conventional conjugate beamforming, where the precoding factor is equal to $\hat{g}^*_{mk}$. Compared to the conventional scheme, the normalized conjugate beamforming just performs a phase-shift of the data signal. 

In \eqref{eq:DLdata}, $\{\eta_{mk}\}$ represent power control coefficients chosen to satisfy the transmit power constraint at the $m$th AP 
\begin{equation}
\label{eq:pwConstraint}
\mathbb{E}\{|x_m|^2\}\leq\rho_\textrm{d}.
\end{equation}
Plugging \eqref{eq:DLdata} into \eqref{eq:pwConstraint}, we obtain
\begin{equation}
\label{eq:STpwC}
\sum \limits_{k=1}^K \eta_{mk} \leq 1, \ \text{for all}\ m,
\end{equation}  
which represents the short-term average power constraint.\footnote{In contrast, for conventional conjugate beamforming the corresponding long-term average power constraint is given, in \cite{NgoCF}, by $\sum_{k=1}^K \mu_{mk}\gamma_{mk} \leq 1, \ \text{for all}\ m=1,\ldots,M$, where $\{\mu_{mk}\}$ are the power control coefficients.}
\begin{figure*}[!t]
\normalsize
\setcounter{MYtempeqncnt}{\value{equation}}
\setcounter{equation}{\ccounter}
\begin{gather}
R^{\textrm{st}}_k \approx \log_2 \left(1+\frac{\rho_{\textrm{d}}\frac{\pi}{4}\left(\sum\limits^M_{m=1}\sqrt{\eta_{mk}\gamma_{mk}}\right)^2}{\rho_{\textrm{d}}\frac{4}{\pi}\sum\limits^K_{k' \neq k} \sum\limits^M_{m=1} \sum\limits^M_{n \neq m} \frac{\sqrt{\eta_{mk'}\eta_{nk'}\gamma_{mk'}\gamma_{nk'}}\beta_{mk}\beta_{nk}}{\beta_{mk'}\beta_{nk'}} \left|\bm{\varphi}^H_{k'} \bm{\varphi}_{k}\right|^2+\rho_{\textrm{d}}\sum\limits^K_{k'=1}\sum\limits^M_{m=1}\eta_{mk'}\beta_{mk} - \rho_{\textrm{d}}\frac{\pi}{4}\sum\limits^M_{m=1}\eta_{mk}\gamma_{mk} + 1} \right), \label{eq:RateST} \\
R^{\textrm{lt}}_k = \log_2 \left( 1 + \frac{\rho_{\textrm{d}}\left(\sum\limits^M_{m=1} \sqrt{\mu_{mk}}\gamma_{mk}\right)^2}{\rho_{\textrm{d}}\sum\limits^K_{k' \neq k}\left(\sum\limits^M_{m=1} \sqrt{\mu_{mk'}}\gamma_{mk'}\frac{\beta_{mk}}{\beta_{mk'}}\right)^2 |\bm{\varphi}^H_{k'}\bm{\varphi}_k|^2 + \rho_{\textrm{d}} \sum\limits^K_{k'=1} \sum\limits^M_{m=1} \mu_{mk'} \beta_{mk} \gamma_{mk'} + 1}\right), \label{eq:RateLT}
\end{gather}
\setcounter{equation}{\value{MYtempeqncnt}}
\hrulefill
\vspace*{4pt}
\end{figure*}

The signal received at the $k$th user is
\begin{align}
\label{eq:signalUE1}
r_{\textrm{d},k} &= \sum^M_{m=1} g_{mk} x_m + w_{\textrm{d},k} \nonumber \\ 
&= \sqrt{\rho_{\textrm{d}}} \sum \limits_{m=1}^M \sum \limits^K_{k'=1} \sqrt{\eta_{mk'}} {g}_{mk} \frac{\hat{g}^*_{mk'}}{|\hat{g}_{mk'}|} q_{k'} + w_{\textrm{d},k},  
\end{align} 
where $ w_{\textrm{d},k} $ is additive $\mathcal{CN}(0,1)$ noise. User $k$ then detects $q_k$ from $r_{\textrm{d},k}$.

\section{Performance Analysis}

In this section we derive an achievable downlink rate as well as relevant performance indicators like \textit{beamforming gain uncertainty} (BU), \textit{user interference} (UI), and strength of the \textit{desired signal} (DS). A tight approximation of the achievable rate is proposed and computed in closed form.

By definition, an achievable downlink rate for a user is the mutual information between the observed signal, the partial knowledge of the channel and the unknown transmitted signal. Since no downlink training is performed, the users use only statistical properties of the channels to decode the downlink data. From \eqref{eq:signalUE1}, the signal received by the $k$th user can be rewritten as
\begin{align}
\label{eq:signalUE}
r_{\textrm{d},k} 
&= \text{DS}_k \cdot q_k + \text{BU}_k \cdot q_k + \sum \limits^K_{k'\neq k}\text{UI}_{kk'} \cdot q_{k'} + w_{\textrm{d},k},  
\end{align}  
where $\text{DS}_k$, $\text{BU}_k$, and $\text{UI}_{kk'}$ reflect the coherent beamforming gain, beamforming gain uncertainty, and inter-user interference, respectively, given by 
\begin{align}
& \text{DS}_k = \sqrt{\rho_{\textrm{d}}} \ \mathbb{E}\left\{\sum^M_{m=1} \sqrt{\eta_{mk}} g_{mk} \frac{\hat{g}^*_{mk}}{|\hat{g}_{mk}|}\right\}, \label{eq:DS}\\
& \text{BU}_k = \sqrt{\rho_{\textrm{d}}} \left(\sum^M_{m=1} \sqrt{\eta_{mk}} g_{mk} \frac{\hat{g}^*_{mk}}{|\hat{g}_{mk}|} \right. \nonumber \\ 
& \qquad\qquad\qquad\qquad \left. - \mathbb{E}\left\{\sum^M_{m=1} \sqrt{\eta_{mk}} g_{mk} \frac{\hat{g}^*_{mk}}{|\hat{g}_{mk}|}\right\}\right), \label{eq:BU}\\
& \text{UI}_{kk'} = \sqrt{\rho_{\textrm{d}}} \sum^M_{m=1} \sqrt{\eta_{mk'}} g_{mk} \frac{\hat{g}^*_{mk'}}{|\hat{g}_{mk'}|}.
\label{eq:UI}
\end{align}  
Since $q_k$ is independent of $\text{BU}_k$, the first and the second terms in (\ref{eq:signalUE}) are uncorrelated. Furthermore, since $q_k$ and $q_{k'}$ are independent, the first term is uncorrelated with the third term as well as the noise (fourth) term by assumption. Therefore, the sum of the second, third, and fourth terms in (\ref{eq:signalUE}) can be treated as an uncorrelated effective  noise. Recalling that uncorrelated Gaussian noise yields a capacity lower bound \cite{Medard}, we can obtain the following capacity lower bound (achievable downlink rate) for the $k$th user:
\begin{equation}
\label{eq:RateDL}
R^{\textrm{st}}_k = \log_2 \left( 1 + \frac{|\text{DS}_k|^2}{\mathbb{E}\{|\text{BU}_k|^2\}+\sum\limits^K_{k' \neq k}\mathbb{E}\{|\text{UI}_{kk'}|^2\}+1} \right).
\end{equation}  
Terms $\text{DS}_k$ and $\mathbb{E}\{|\text{BU}_k|^2\}$ in \eqref{eq:RateDL} can be exactly computed in closed form. However, it is very difficult to obtain a closed form result for term  $\mathbb{E}\{|\text{UI}_{kk'}|^2\}$  due to analytically intractable form of $\mathbb{E}\left\{g_{mk}\frac{\hat{g}^*_{mk'}}{\left|\hat{g}^*_{mk'}\right|} \right\}$. To alleviate
this difficulty, we use the first order approximation as in \cite{YuYiu}:
\begin{equation}\label{eq:approx}
\mathbb{E}\left\{g_{mk}\frac{\hat{g}^*_{mk'}}{\left|\hat{g}^*_{mk'}\right|} \right\} \approx \frac{\mathbb{E}\{g_{mk}\hat{g}^*_{mk'}\}}{\mathbb{E}\{\left|\hat{g}^*_{mk'}\right| \}}.
\end{equation}
In Section~\ref{sec:numerical-results}, we will numerically show that this approximation is very accurate. By using \eqref{eq:approx}, we obtain an approximate closed-form expression for the achievable rate given by \eqref{eq:RateDL} as follows.

\textit{Proposition 1}: With normalized conjugate beamforming, the achievable rate of the transmission from the APs to the $k$th user given in \eqref{eq:RateDL} can be approximated by  (\ref{eq:RateST}) shown at the top of the page.
\begin{IEEEproof}
See Appendix A.

\

\end{IEEEproof}
\begin{table*}[!t]
\caption{normalized vs conventional conjugate beamforming scheme}
\label{tab:formulas}
\centering
\renewcommand{\arraystretch}{2}
\begin{tabular}{l | c}
\hline
Normalized conj. beamforming $|\text{DS}_k|^2$ & $\rho_{\textrm{d}}\frac{\pi}{4}\left(\sum^M_{m=1}\sqrt{\eta_{mk}\gamma_{mk}}\right)^2$ \\ 
Conventional conj. beamforming $|\text{DS}_k|^2$ & $\rho_{\textrm{d}}\left(\sum^M_{m=1} \sqrt{\mu_{mk}}\gamma_{mk}\right)^2$\\ \hline 
Normalized conj. beamforming $\mathbb{E}\{|\text{BU}_k|^2\}$ & $\rho_{\textrm{d}} \sum^M_{m=1}\eta_{mk}\left(\beta_{mk}-\frac{\pi}{4}\gamma_{mk}\right)$ \\ 
Conventional conj. beamforming $\mathbb{E}\{|\text{BU}_k|^2\}$ & $\rho_{\textrm{d}} \sum^M_{m=1}\mu_{mk}\beta_{mk}\gamma_{mk}$\\ \hline 
Normalized conj. beamforming $\mathbb{E}\{|\text{UI}_{kk'}|^2\}$ & $\rho_{\textrm{d}}\frac{4}{\pi}\left|\bm{\varphi}^H_{k'} \bm{\varphi}_{k}\right|^2\sum\limits^M_{m=1} \sum\limits^M_{n \neq m} \frac{\sqrt{\eta_{mk'}\eta_{nk'}\gamma_{mk'}\gamma_{nk'}}\beta_{mk}\beta_{nk}}{\beta_{mk'}\beta_{nk'}}+\rho_{\textrm{d}}\sum\limits^M_{m=1}\eta_{mk'}\beta_{mk}$ \\ 
Conventional conj. beamforming $\mathbb{E}\{|\text{UI}_{kk'}|^2\}$ & $\rho_{\textrm{d}}\left|\bm{\varphi}^H_{k'}\bm{\varphi}_k\right|^2 \left(\sum\limits^M_{m=1} \sqrt{\mu_{mk'}}\gamma_{mk'}\frac{\beta_{mk}}{\beta_{mk'}}\right)^2 + \rho_{\textrm{d}} \sum\limits^M_{m=1} \mu_{mk'} \beta_{mk} \gamma_{mk'}$ \\ \hline
\end{tabular}

\

\hrulefill
\vspace*{4pt}
\end{table*}

The achievable downlink rate for the conventional conjugate beamforming is given, in \cite{NgoCF}, by (\ref{eq:RateLT}) shown at the top of the page. Note that, the superscripts \{st, lt\} stand for short-term, and long-term, respectively, in order to emphasize the relationship between precoding scheme and power constraint. In (\ref{eq:RateLT}), $\{\mu_{mk}\}$ define the power control coefficients for the conventional conjugate beamforming scheme. 
Table~\ref{tab:formulas} analytically compares both the precoding schemes by giving the explicit formulas of each single term forming the effective signal-to-interference-plus-noise ratio (SINR) in (\ref{eq:RateDL}).

\section{Numerical Results}
\label{sec:numerical-results}
Next, we numerically evaluate the impact of the short-term average power constraint given by (\ref{eq:STpwC}) on the achievable downlink rate, coherent beamforming gain, beamforming gain uncertainty, and user interference due to the channel non-orthogonality, in a cell-free massive MIMO system. Then, we compare the performance provided by normalized and conventional conjugate beamforming  precoding schemes.

\subsection{Simulation Scenario and Parameters}        
We consider a cell-free massive MIMO system with $M$ APs and $K$ users uniformly spread out at random within a nominal squared area of size $1\times 1$ km$^2$. In order to simulate a network with an infinite area and without boundaries, we implement a wrap-around technique, in which the nominal area is wrapped around by eight neighbor replicas.

The large-scale fading coefficient $\beta_{mk}$ is modeled as
\setcounter{equation}{18}
\begin{equation}
\label{eq:beta}
\beta_{mk} = \text{PL}_{mk} \cdot 10^{\frac{\sigma_{sh}z_{mk}}{10}}
\end{equation}  
where $\text{PL}_{mk}$ describes the path loss, and $10^{\frac{\sigma_{sh}z_{mk}}{10}}$ is the shadowing with standard deviation $\sigma_{sh}$ and $z_{mk}\sim\mathcal{N}(0,1)$. 

For all examples, we adopt the three-slope model for the path loss as defined in \cite{NgoCF} and uncorrelated shadow fading. We also used the following setup: the carrier frequency is 1.9 GHz, the bandwidth is 20 MHz, and the coherence interval length is $\tau = 200$ symbols. The uplink training duration  is $\tau_{\textrm{u,p}} = K/2$.  The AP and user antenna height is 15 m, 1.65 m, respectively, and the antenna gains are $0$ dBi. The noise figure for both the uplink and the downlink is 9 dB, and the radiated power is 200 mW for APs and 100 mW for users. 

We assume that the random pilot assignment scheme \cite{NgoCF} is used. More precisely, each user is assigned randomly a pilot sequence from a pre-defined pilot set which includes $\tau_{\textrm{u,p}}$ orthogonal pilot sequences of length $\tau_{\textrm{u,p}}$.

We also assume that pilots and data signals are always transmitted with full power, i.e., no power control is performed. To guarantee a fair comparison between normalized conjugate beamforming and conventional conjugate beamforming, we choose the power control coefficients as
\begin{align}
\label{eq:eta} 
\begin{cases}
\eta_{mk} = \frac{\gamma_{mk}}{\sum\nolimits^K_{k'=1}\gamma_{mk'}}, &\text{for normalized conj. beam.,}\\
\mu_{mk} = \frac{1}{\sum\nolimits^K_{k'=1}\gamma_{mk'}}, &\text{for conventional conj. beam.,}\\
\end{cases}
\end{align}
for all $k = 1, \ldots, K$. With the power control coefficients given in \eqref{eq:eta}, the powers spent by the $m$th AP on the $k$th user are the same for both schemes.

\subsection{Performance Evaluation}
\begin{figure}[!t]
\centering
\includegraphics[width=3.3in]{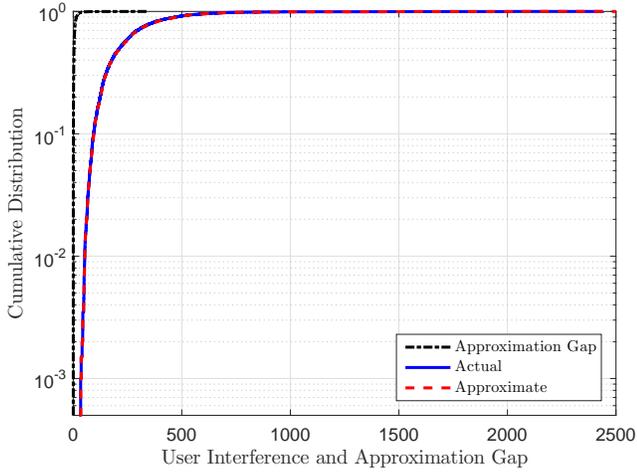}
\caption{The cumulative distribution of user interference term, and the cumulative distribution of the approximation gap between the actual and the approximate values of user interference, for the normalized conjugate beamforming scheme. Here, $M = 100$, $K = 40$.}
\label{fig:approx}
\end{figure}

First, we evaluate the tightness of our approximation \eqref{eq:approx}. Figure~\ref{fig:approx} shows the cumulative distribution of user interference term, and the cumulative distribution of the ``Approximation Gap'', which is the absolute value of the difference between the ``Actual'' and the ``Approximate'' values of user interference term, for  $M = 100$, and $K = 40$. The ``Approximate'' curve is obtained by using the approximate closed-form expression for term $\mathbb{E}\{|\textrm{UI}_{kk'}|^2\}$, shown in Table \ref{tab:formulas}. The ``Actual'' curve is generated by  using Monte Carlo simulation. We can see that our proposed approximation is very accurate with high probability and the approximation gap is very small compared to user interference value.

\begin{figure}[!t]
\centering
\includegraphics[width=3.3in]{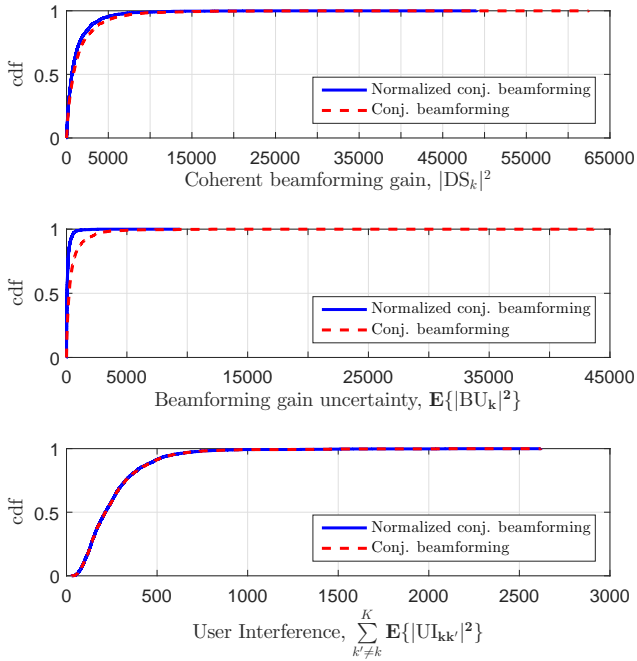}
\caption{The cumulative distribution of the coherent beamforming gain, the beamforming uncertainty gain, and the user interference, for the normalized and  conventional conjugate beamforming scheme. Here, $M = 100$, $K = 40$.}
\label{fig:bfterms}
\end{figure}

Next, we compare the performances of the normalized and the conventional conjugate beamforming in terms of: \textit{(i)} coherent beamforming gain; \textit{(ii)} beamforming gain uncertainty; and \textit{(iii)} user interference due to the channel non-orthogonality, by using the formulas listed in Table \ref{tab:formulas}. Figure~\ref{fig:bfterms} illustrates the cumulative distribution of these three performance factors, for both the precoding schemes at $M = 100$, $K = 40$.      

As we can see from the figure, compared with the conventional conjugate beamforming, the normalized conjugate beamforming provides a lower coherent beamforming gain, but a lower beamforming gain uncertainty. Moreover, both the precoding schemes offer a similar effect of user interference. The coherent beamforming gap is trivially equal to $\pi/4$, corresponding to about 21\% at the median point. The beamforming uncertainty gap is equal to  $\rho_{\textrm{d}}\frac{\pi}{4} \sum^M_{m=1}\eta_{mk}\gamma_{mk}$ which represents about 75\% at the median point. 

\begin{figure}[!t]
\centering
\includegraphics[width=3.3in]{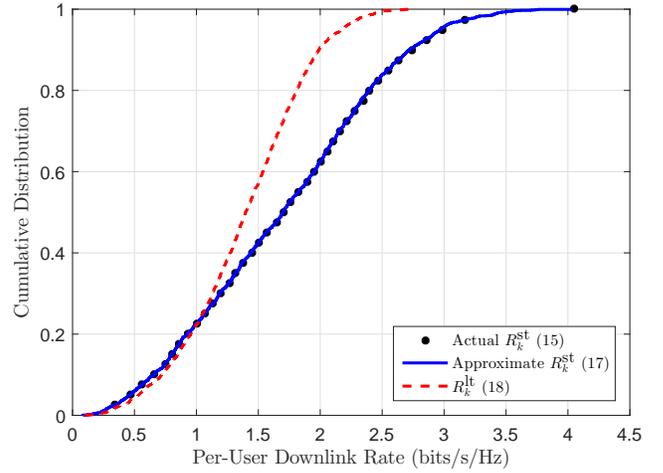}
\caption{The cumulative distribution of the per-user achievable downlink rate for the case of normalized (actual and approximate $R^{\text{st}}_k$) and conventional ($R^{\text{lt}}_k$) conjugate beamforming scheme. Here, $M = 100$, $K = 40$.}
\label{fig:rateM100K40}
\end{figure}

\begin{figure}[!t]
\centering
\includegraphics[width=3.3in]{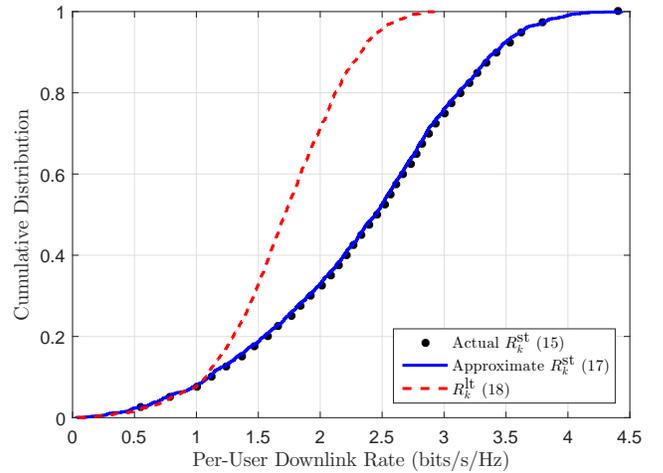}
\caption{The same as Figure~\ref{fig:rateM100K40}, but $M = 50$ and $K = 10$.}
\label{fig:rateM50K10}
\end{figure}

Finally, we compare the performances of the normalized and conventional conjugate beamforming schemes in terms of achievable downlink rates. Figure~\ref{fig:rateM100K40} shows the cumulative distribution of the per-user achievable downlink rates given by (\ref{eq:RateST}) and (\ref{eq:RateLT}), respectively, with  $M = 100$ and $K = 40$. It also emphasizes the accuracy of our approximation by comparing (\ref{eq:RateDL}) and (\ref{eq:RateST}).  
In Figure~\ref{fig:rateM50K10}, we consider the same performance metric, but for a cell-free massive MIMO system with lower density, i.e., $M = 50$ and $K = 10$.  
As we can see, the 95\%-likely per-user achievable downlink rates of the normalized and conventional conjugate beamforming schemes are comparable. However,  
normalized conjugate beamforming  outperforms the conventional one in terms of median per-user achievable downlink rate, especially when $M=50$, as shown in Figure~\ref{fig:rateM50K10}. Reducing the beamforming uncertainty gain is one of the the main advantages of normalized conjugate beamforming scheme over the conventional scheme. When the number of APs is very large, the channel hardens, and hence, the beamforming uncertainty gain is small for both schemes. This is the reason for why there is a modest performance improvement by increasing $M$, as shown in Figure \ref{fig:rateM100K40}.        

\section{Conclusion and Future Work}

In this paper, we analyzed the performance of cell-free massive MIMO systems with the normalized conjugate beamforming scheme. This scheme satisfies a short-term average power constraint at the APs where the average is taken only over the codewords. A tightly approximate closed-form expression for the achievable rate was derived. Compared to the conventional conjugate scheme in \cite{NgoCF}, the normalized conjugate beamforming scheme suffers less beamforming gain uncertainty. As a result, the normalized conjugate beamforming performs better than the conventional conjugate beamforming when the number of APs is moderate.

Further work may include studying the max-min power control and more advanced pilot assignment algorithms. The downlink beamforming training scheme may also be addressed following a similar methodology as in \cite{IntGlobecom16} with respect to conventional conjugate beamforming.

\section*{Appendix}
\subsection{Proof of Proposition 1}
From \eqref{eq:DS}, and by using $g_{mk}\triangleq\hat{g}_{mk}+\tilde{g}_{mk}$ where $\tilde{g}_{mk}$ and $\hat{g}_{mk}$ are independent, we have
\begin{align}
\label{DSproof}
& |\text{DS}_k|^2 = \rho_{\textrm{d}} \left|\mathbb{E}\left\{\sum^M_{m=1} \sqrt{\eta_{mk}} \left(|\hat{g}_{mk}| + \tilde{g}_{mk} \frac{\hat{g}^*_{mk}}{|\hat{g}_{mk}|} \right)\right\}\right|^2 \nonumber \\
& = \rho_{\textrm{d}} \left|\sum^M_{m=1} \sqrt{\eta_{mk}} \ \mathbb{E} \{|\hat{g}_{mk}|\}\right|^2 = \frac{\pi \rho_{\textrm{d}}}{4} \left(\sum^M_{m=1} \sqrt{\eta_{mk}\gamma_{mk}} \right)^2.
\end{align}
In the last equality we have used $\mathbb{E} \{|\hat{g}_{mk}|\} = \frac{\sqrt{\pi}}{2}\sqrt{\gamma_{mk}}$, since the magnitude of a complex normal RV follows a Rayleigh distribution. 

Similarly, from \eqref{eq:BU} we have  
\begin{align}
\label{BUproof}
&\mathbb{E}\left\{\left|\text{BU}_k\right|^2\right\} = \rho_{\textrm{d}} \mathrm{Var}\left\{\sum^M_{m=1} \sqrt{\eta_{mk}} \left(|\hat{g}_{mk}| + \tilde{g}_{mk} \frac{\hat{g}^*_{mk}}{|\hat{g}_{mk}|} \right)\right\} \nonumber \\
& = \rho_{\textrm{d}} \sum^M_{m=1} \eta_{mk} \left(\mathbb{E}\left\{\left| |\hat{g}_{mk}| + \tilde{g}_{mk} \frac{\hat{g}^*_{mk}}{|\hat{g}_{mk}|} \right|^2\right\} \right. \nonumber \\
&\quad\quad \left. - \left|\mathbb{E}\left\{|\hat{g}_{mk}| + \tilde{g}_{mk} \frac{\hat{g}^*_{mk}}{|\hat{g}_{mk}|} \right\}\right|^2 \right) \nonumber \\
& = \rho_{\textrm{d}} \sum^M_{m=1} \eta_{mk} \left( \mathbb{E}\left\{|\hat{g}_{mk}|^2 \right\} + \mathbb{E}\left\{|\tilde{g}_{mk}|^2  \right\} - \left| \mathbb{E}\left\{|\hat{g}_{mk} |\right\} \right|^2 \right) \nonumber \\
& = \rho_{\textrm{d}} \sum^M_{m=1}\eta_{mk}\left(\beta_{mk}-\frac{\pi}{4}\gamma_{mk}\right).
\end{align}

According to \eqref{eq:UI}, we have
\begin{align}
\label{UIproof}
& \mathbb{E}\left\{\left|\text{UI}_{kk'}\right|^2\right\} = \rho_{\textrm{d}} \mathbb{E}\left\{\left|\sum^M_{m=1} \sqrt{\eta_{mk'}} g_{mk} \frac{\hat{g}^*_{mk'}}{|\hat{g}_{mk'}|}\right|^2 \right\} \nonumber \\
& = \rho_{\textrm{d}} \mathbb{E} \left\{ \sum^M_{m=1} \sum^M_{n=1} \sqrt{\eta_{mk'} \eta_{nk'}}g_{mk} g^*_{nk} \frac{\hat{g}^*_{mk'}}{|\hat{g}_{mk'}|} \frac{\hat{g}_{nk'}}{|\hat{g}_{nk'}|} \right\} \nonumber \\
& = \rho_{\textrm{d}} \sum^M_{m=1} \sum^M_{n \neq m} \sqrt{\eta_{mk'} \eta_{nk'}} \mathbb{E} \left\{g_{mk} \frac{\hat{g}^*_{mk'}}{|\hat{g}_{mk'}|} \right\} \mathbb{E}\left\{g^*_{nk}  \frac{\hat{g}_{nk'}}{|\hat{g}_{nk'}|} \right\} \nonumber \\
&\quad + \rho_{\textrm{d}} \sum^M_{m=1} \eta_{mk'} \mathbb{E} \left\{ |g_{mk}|^2 \right\}.
\end{align}
Now, we apply approximation \eqref{eq:approx} to \eqref{UIproof}. First, we derive
\begin{align}
\label{eq:approxproof}
& \mathbb{E} \{ g_{mk} \hat{g}^*_{mk'} \} = c_{mk'} \Bigg(\sqrt{\tau_{\textrm{u,p}}\rho_{\textrm{u,p}}} \left(\bm{\varphi}^H_{k'} \bm{\varphi}_k\right)^\ast \mathbb{E}\{ |g_{mk}|^2 \} \Bigg. \nonumber \\
&\quad + \Bigg. \sqrt{\tau_{\textrm{u,p}}\rho_{\textrm{u,p}}} \sum\limits^K_{i \neq k} \bm{\varphi}^T_{k'} \bm{\varphi}_i^\ast \mathbb{E} \{ g_{mk} g^*_{mi} \} + \mathbb{E}\{g_{mk} \bm{\varphi}^T_{k'}\textbf{w}_{\textrm{up},m}^\ast \} \Bigg) \nonumber \\
& = c_{mk'} \sqrt{\tau_{\textrm{u,p}}\rho_{\textrm{u,p}}} \bm{\varphi}^T_{k'} \bm{\varphi}_k^\ast \beta_{mk},
\end{align}
since $g_{mk}$ is a zero-mean RV independent of $g_{mi}$ and $\textbf{w}_{\textrm{up},m}$. Plugging \eqref{eq:approxproof} into \eqref{eq:approx}, and in turn \eqref{eq:approx} in \eqref{UIproof}, we get
\begin{align}
\label{UIproof2}
& \mathbb{E}\left\{\left|\text{UI}_{kk'}\right|^2\right\} = \rho_{\textrm{d}} \sum^M_{m=1} \eta_{mk'} \beta_{mk} \nonumber \\
& + \frac{4\rho_{\textrm{d}}}{\pi}\tau_{\textrm{u,p}}\rho_{\textrm{u,p}}|\bm{\varphi}^H_{k'} \bm{\varphi}_k|^2 \sum^M_{m=1} \sum^M_{n \neq m} \frac{\sqrt{\eta_{mk'}\eta_{nk'}}c_{mk'}c_{nk'}\beta_{mk}\beta_{nk}}{\sqrt{\gamma_{mk'}\gamma_{nk'}}} \nonumber \\
& = \rho_{\textrm{d}}\sum\limits^M_{m=1}\eta_{mk'}\beta_{mk} \nonumber \\
& + \rho_{\textrm{d}}\frac{4}{\pi}\left|\bm{\varphi}^H_{k'} \bm{\varphi}_{k}\right|^2\sum\limits^M_{m=1} \sum\limits^M_{n \neq m} \frac{\sqrt{\eta_{mk'}\eta_{nk'}\gamma_{mk'}\gamma_{nk'}}\beta_{mk}\beta_{nk}}{\beta_{mk'}\beta_{nk'}},
\end{align} 
where in the last equality, we have used \eqref{eq:defGamma}.
Lastly, by substituting \eqref{DSproof}, \eqref{BUproof}, and \eqref{UIproof2} in \eqref{eq:RateDL}, we get \eqref{eq:RateST}.   

\bibliographystyle{IEEEtran}
\bibliography{IEEEabrv,mybib}

\begin{thebibliography}{10}
\providecommand{\url}[1]{#1}
\csname url@samestyle\endcsname
\providecommand{\newblock}{\relax}
\providecommand{\bibinfo}[2]{#2}
\providecommand{\BIBentrySTDinterwordspacing}{\spaceskip=0pt\relax}
\providecommand{\BIBentryALTinterwordstretchfactor}{4}
\providecommand{\BIBentryALTinterwordspacing}{\spaceskip=\fontdimen2\font plus
\BIBentryALTinterwordstretchfactor\fontdimen3\font minus
  \fontdimen4\font\relax}
\providecommand{\BIBforeignlanguage}[2]{{%
\expandafter\ifx\csname l@#1\endcsname\relax
\typeout{** WARNING: IEEEtran.bst: No hyphenation pattern has been}%
\typeout{** loaded for the language `#1'. Using the pattern for}%
\typeout{** the default language instead.}%
\else
\language=\csname l@#1\endcsname
\fi
#2}}
\providecommand{\BIBdecl}{\relax}
\BIBdecl

\bibitem{NgoCF}
\BIBentryALTinterwordspacing
H.~Q. Ngo, A.~E. Ashikhmin, H.~Yang, E.~G. Larsson, and T.~L. Marzetta,
  ``Cell-free massive {MIMO} versus small cells,'' \emph{{IEEE} Trans. Wireless
  Commun.}, 2016, submitted. [Online]. Available:
  \url{https://arxiv.org/abs/1602.08232}
\BIBentrySTDinterwordspacing

\bibitem{ZhouWCS}
S.~Zhou, M.~Zhao, X.~Xu, J.~Wang, and Y.~Yao, ``Distributed wireless
  communication system: a new architecture for future public wireless access,''
  \emph{{IEEE} Commun. Mag.}, vol.~41, no.~3, pp. 108--113, Mar. 2003.

\bibitem{Truong}
K.~T. Truong and R.~W. Heath, ``The viability of distributed antennas for
  massive {MIMO} systems,'' in \emph{Proc. Asilomar Conference on Signals,
  Systems and Computers}, Nov. 2013, pp. 1318--1323.

\bibitem{MarzettaNonCooperative}
T.~L. Marzetta, ``Noncooperative cellular wireless with unlimited numbers of
  base station antennas,'' \emph{{IEEE} Trans. Wireless Commun.}, vol.~9,
  no.~11, pp. 3590--3600, Nov. 2010.

\bibitem{Kamga}
G.~N. Kamga, M.~Xia, and S.~Aïssa, ``Spectral-efficiency analysis of massive
  {MIMO} systems in centralized and distributed schemes,'' \emph{{IEEE} Trans.
  Commun.}, vol.~64, no.~5, pp. 1930--1941, May 2016.

\bibitem{Khoshnevisan}
M.~Khoshnevisan and J.~N. Laneman, ``Power allocation in multi-antenna wireless
  systems subject to simultaneous power constraints,'' \emph{{IEEE} Trans.
  Commun.}, vol.~60, no.~12, pp. 3855--3864, Dec. 2012.

\bibitem{Lund}
J.~Vieira, F.~Rusek, and F.~Tufvesson, ``Reciprocity calibration methods for
  massive {MIMO} based on antenna coupling,'' in \emph{Proc. {IEEE} Global
  Communications Conference (GLOBECOM)}, Dec. 2014, pp. 3708--3712.

\bibitem{PilotContamination}
J.~Jose, A.~Ashikhmin, T.~L. Marzetta, and S.~Vishwanath, ``Pilot contamination
  and precoding in multi-cell {TDD} systems,'' \emph{{IEEE} Trans. Wireless
  Commun.}, vol.~10, no.~8, pp. 2640--2651, Aug. 2011.

\bibitem{Medard}
M.~Medard, ``The effect upon channel capacity in wireless communications of
  perfect and imperfect knowledge of the channel,'' \emph{{IEEE} Trans. Inf.
  Theory}, vol.~46, no.~3, pp. 933--946, May 2000.

\bibitem{YuYiu}
L.~Yu, W.~Liu, and R.~Langley, ``{SINR} analysis of the subtraction-based {SMI}
  beamformer,'' \emph{{IEEE} Trans. Signal Process.}, vol.~58, no.~11, pp.
  5926--5932, Nov. 2010.

\bibitem{IntGlobecom16}
\BIBentryALTinterwordspacing
G.~Interdonato, H.~Ngo, E.~G. Larsson, and P.~Frenger, ``How much do downlink
  pilots improve {Cell-Free} massive {MIMO?}'' in \emph{Proc. {IEEE} Global
  Communications Conference ({GLOBECOM})}, Dec. 2016, accepted. [Online].
  Available: \url{http://arxiv.org/abs/1607.04753}
\BIBentrySTDinterwordspacing

\end{thebibliography}

\end{document}